\documentclass{mn2e}
\usepackage{epsf}

\title[Accretion disc viscosity: how big is alpha?]
{Accretion disc viscosity: how big is alpha?}

\author[A. R. King, J. E. Pringle \& M. Livio] {A. R. King$^1$,
J.E. Pringle$^{1,2, 3}$ and M. Livio$^3$\\ $^1$Theoretical
Astrophysics Group, University of Leicester, Leicester LE1 7RH\\
$^2$Institute of Astronomy, University of Cambridge, Madingley Road,
Cambridge CB3 0HA\\
$^3$Space Telescope Science Institute, 3700 San
Martin Drive, Baltimore, MD 21218, USA}

\date{\today}

\volume{000}

\pagerange{\pageref{firstpage}--\pageref{lastpage}} \pubyear{2006}

\begin{document}

\label{firstpage}

\maketitle

\begin{abstract}

We consider observational and theoretical estimates of the accretion
disc viscosity parameter $\alpha$. We find that in thin, fully-ionized
discs, the best observational evidence suggests a typical range
$\alpha \sim 0.1 - 0.4$, whereas the relevant numerical simulations
tend to derive estimates for $\alpha$ which are an order of magnitude
smaller. We discuss possible reasons for this apparent discrepancy.

\end{abstract}


\section{Introduction}

Accretion discs are believed to be present in a wide variety of
astronomical systems, and have been a major research topic for several
decades (see e.g. Pringle, 1981 and Frank, King \& Raine, 2002). For
much of this time theorists have had problems understanding the
fundamental driving mechanism transporting angular momentum outwards
and thus allowing matter to spiral inwards in a disc. This mechanism
is usually called `viscosity', and appears in virtually all of disc
theory. Ideas about discs nevertheless gained credibility because of
two factors. The first was that the radial distribution of effective
temperature across a steady disc ($T(R) \propto R^{-3/4}$) is
independent of the viscosity, being just a statement of energy
conservation, and is in reasonable accord with both continuum spectra
and eclipse mapping of cataclysmic variables (CVs). These are close
binary systems where a white dwarf accretes from a low--mass companion
via a disc. The second was the devising of a physically--motivated,
dimensionless scaling of the kinematic viscosity $\nu$ as
\begin{equation}
\nu = \alpha c_s H
\label{alpha}
\end{equation}
(Shakura \& Sunyaev, 1973). Here $c_s$ is the local mean sound speed
in the disc, and $H \sim (c_s/v_{\phi})R$ is the scaleheight
perpendicular to the disc plane at radius $R$, where $v_{\phi}$ is the
azimuthal velocity.  In a thin disc (i.e. one with cooling efficient
enough that $H \ll R$) where the velocity $v_{\phi}$ is very close to
the Kepler value $v_K = (GM/R)^{1/2}$ (where $M$ is the accreting
central mass), these quantities are well--defined, so $\alpha$ is a
dimensionless quantity specifying the local rate at which angular
momentum (strictly speaking, that component orthogonal to the disc
plane) is transported. The parameter $\alpha$ is a quantity whose
properties are to be determined experimentally.~\footnote{It is
important to realise that $\alpha$ is a mean quantity, averaged
perpendicular to the disc plane. It is well-defined only if the disc
is thin, i.e. $H \ll R$. In a thick disc, with $H \sim R$, $\alpha$
has no clear meaning}

This alpha--prescription allows formal closure of the system of
equations describing a thin disc, even though there is no presumption
that $\alpha$ is anything other than an unknown dimensionless scaling
variable, although it is often assumed to be a constant. Gratifyingly,
many of the properties of steady thin discs turn out to have rather
weak dependences on $\alpha$. Ignorance of the physical properties or
strength of the angular momentum transport process, represented in
dimensionless fashion by $\alpha$, is thus much less of an obstacle to
practical application of this simple picture than one might
suppose. The perceived success of these applications amounts to noting
that fairly similar values of $\alpha$ appear to give reasonable
agreement with observations of many systems.

A physically plausible theory of the underlying causes of disc
`viscosity' has emerged over the last fifteen years. In their seminal
paper Shakura \& Sunyaev (1973) argued that magnetic fields are the
likely way in which a shearing disc flow transports angular momentum
from rapidly--rotating fluid to more slowly--rotating fluid further
out.  This concept was given impetus with the realization (Balbus \&
Hawley, 1991) that what is now called the magnetorotational
instability (MRI) can provide the necessary feedback to maintain a
magnetic dynamo in accretion discs. The MRI forms the basis of all
current theoretical simulations of this process. These have not yet
reached the point where direct comparison with observation is
possible, in terms for example of being able to predict the spectrum
of radiation emitted by accretion discs. Instead the main reason for
optimism has been the belief that these simulations do demonstrate the
feasibility of a self-maintaining process which transports angular
momentum in the required manner.

The point we wish to address in this paper is that it is also often
assumed that numerical simulations produce formal values of $\alpha$
which agree with those inferred from observation. The purpose of this
paper is to examine how far this is true. In Section 2 we discuss
those astronomical phenomena which give the strongest observational
evidence for the value of $\alpha$. These are time-dependent discs
involved in the outbursts of dwarf novae and of the X-ray
transients. The discs in these systems are fully ionized, because of
the nature of the outburst mechanism, and so correspond closely to the
the majority of the numerical simulations which consider full MHD. For
cooler discs which are not sufficiently ionized so that the magnetic
field is not tied strongly enough to the disc gas, the MRI is likely
to be less vigorous or even non-operative, so reducing the expected
value of $\alpha$ (Gammie, 1996). In Section 3 we discuss the
estimates of the value of $\alpha$ derived from numerical
simulations. We concentrate on those simulations which do not impose
an external seed field threading the whole disc to drive the MRI, as
neither in the dwarf novae nor in the X-ray transients is there a
plausible source for such a global field. We note that the most recent
computations obtain values of $\alpha$ smaller than those required by
observations by at least an order of magnitude, and often more. In
Section 4 we discuss the limitations under which numerical simulations
of accretion discs have to operate, driven both by the speed of
current computers and the nature of the numerical algorithms,
especially the boundary conditions. We point out that most of the
limitations are likely to act in the direction of reducing
$\alpha$. In Section 5 we discuss the possibility that the fields
generated by accretion discs are more global than can be easily
accommodated in current numerical simulations, together with the
possible consequences of, and complications arising from, the presence
such global fields.

\section{Observed constraints on alpha}

Since $\alpha$ is a dimensionless measure of the
viscosity\footnote{Note that $\alpha$ is not a measure of an isotropic
viscosity as appears, for example, in the Navier-Stokes equation, but
is, rather, strictly only a measure of the vertically averaged ratio
between the $(R, \phi)$-components of the stress and the rate of
strain tensors.} its properties need to be determined from
observations.  We begin therefore by considering observationally
determined estimates of $\alpha$.  Since as we remarked above, steady
thin disc theory has only a fairly weak dependence on $\alpha$, by far
the most reliable and direct way of estimating this quantity is to
consider time--dependent disc behaviour. The size of $\alpha$ is
directly proportional to the rate at which angular momentum is
transported within the disc, and so is directly related to the
timescale on which a disc can evolve. We note that even so it is not
possible to determine detailed properties of $\alpha$, for example how
$\alpha$ might depend on radius, or $H/R$. Thus the observed values of
$\alpha$ correspond to some appropriate mean value over the disc being
modeled, presumably weighted towards larger radii where the viscous
timescales are longest.

\subsection{Dwarf nova outbursts}

The largest body of evidence here comes from the light curves of dwarf
novae, which are a subclass of cataclysmic variables (CVs) which
undergo outbursts at irregular intervals (Warner, 2003). There is now
general agreement that these outbursts result from the presence of
ionization zones within the disc, allowing this to switch between a
cool, low-ionization, low--viscosity state, and a hot, highly ionized,
high--viscosity state (see Lasota, 2001 for a recent review). In the
hot state the disc evolves on the viscous timescale
\begin{equation}
t_{\rm visc} \sim {R^2\over \nu}
\label{tvisc}
\end{equation}
for a time, before a cooling front propagates through the disc and
returns it to the cool state. The initial slow decay of the outburst
thus allows estimates of $\alpha$ in this hot state. Estimates of
$\alpha$ can therefore be obtained from theoretical modelling of the
outburst lightcurves. Since the disc sizes are known from the system
properties, and since the disc temperatures are known from the spectra
thus determining $H/R$, observation of the evolution timescale of the
outbursts gives a reasonably well-determined estimate of the viscous
timescale and hence of $\alpha$.

Smak (1999) considers the observed relation (Bailey, 1975) between the
decay rate and the orbital period. Since the latter largely fixes the
orbital separation and thus the disc size $R$, the decay rate measures
$\alpha$ directly through (\ref{tvisc}). Other estimates (Smak, 1998,
1999) use the delay between the peak of the outburst in the optical
and the UV, which results from the disc closing a small central hole
around the white dwarf on a viscous timescale $t_{\rm visc}$. These
estimates come from the collective properties of a large sample of
dwarf novae. Other estimates use detailed observations of the outburst
light curves of individual systems (Schreiber et al., 2003, 2004 [SS
Cyg and VW Hyi]; Cannizzo 2001a [VW Hyi, U Gem and SS Cyg]; Cannizzo
2001b [WZ Sge]; Buat--M\'enard et al., 2001 [Z Cam]).  All of these
papers agree that $\alpha$ must lie in a fairly narrow range $\alpha
\simeq 0.1 - 0.3$.

\subsection{Outbursts of X-ray transients}

A second class of accreting binaries that have outbursts is that of
the soft X--ray transients (SXTs) in which the accretor is a black
hole or neutron star rather than a white dwarf as in CVs (Lewin \& van
der Klis, 2006). Even for similar orbital parameters, SXT outbursts
are considerably longer than those of dwarf novae (months rather than
days), and have a different lightcurve shape (exponential for short
orbital periods). This at first presented a challenge to theory. There
were initial attempts to explain this by devising different {\it ad
hoc} forms of viscosity (for example setting $\alpha$ to be a function
of $H/R$; Cannizzo, Chen \& Livio, 1995).  However observations show
that the discs in SXTs (and indeed in all low--mass X--ray binaries)
are optically much brighter than expected on the basis of the
accretion rate revealed by their X--ray luminosities. This extra light
can be directly attributed to irradiation of the outer parts of the
disc by some of the central X--rays (van Paradijs \& McClintock,
1994). King \& Ritter (1998) pointed out that this would force most of
the disc to remain in the hot, high--viscosity state until a
significant fraction of the disc mass had accreted on to the central
black hole/neutron star. This explained the longer duration of SXT
outbursts compared with dwarf novae, and indeed their exponential
shape at short orbital periods (where the whole of the disc can be
efficiently irradiated). Under the assumption of efficient irradiation
Dubus et al. (2001) made detailed models of complete SXT light curves
and found $\alpha \simeq 0.2 - 0.4$.

\subsection{Other systems which  yield estimates of $\alpha$}

\subsubsection{Variability in AGN}

A somewhat more indirect method of estimating $\alpha$ is suggested by
Starling et al. (2004) who look at the optical variability of active
galactic nuclei (AGN) on timescales of months to years. Starling et
al. (2004) measure a two--folding timescale, defined as the timescale
over which the optical luminosity changes by a factor of two. They assume
that the optical emission is generated in a standard, fully ionized,
thin disc, and that the two--folding timescale is given by disc's
local thermal timescale (Pringle, 1981)
\begin{equation}
t_{\rm th} \sim {1\over \alpha}\biggl({R\over v_K}\biggr).
\label{thermal}
\end{equation}
They find that $0.01 \leq \alpha \leq 0.03$ for $0.1 \leq L/L_E \leq
1$. Starling et al. note that these values of $\alpha$ are really
lower limits because data sampling means that they might miss shorter
timescales.

\subsubsection{Protostellar accretion discs}

Young pre-main sequence stars are often surrounded by accretion discs
(e.g. Hartmann, 1998). Estimates of the lifetimes of these discs can
be obtained by comparing disc frequencies among stars of different
ages.  Estimates for $\alpha$ in protostellar (T Tauri) discs, based
on evolutionary lifetimes, are given by Hartmann et al. (1998). They
give estimates of $\alpha \approx 0.01$ at disc radii $R \sim 10 -
100$ AU. All of the examples discussed in Sections 2.1 and 2.2 involve
accretion discs which are sufficiently hot that they are fully
ionized. However, at such large radii the protostellar discs are cool
enough that they are unlikely to be fully ionized. If the ionization
fraction is sufficiently low the numerical MHD simulations are not
strictly applicable and the MRI which is thought to drive the
viscosity mechanism is significantly suppressed (Gammie, 1996).

\subsubsection{FU Orionis outbursts}

Models for the outbursts of the pre-main sequence FU Orionis stars in
terms of thermally driven disc outbursts of the kind seen in CVs and
SXTs are given by Clarke et al. (1990), Bell \& Lin (1994), and Lodato
\& Clarke (2004). In order to fit the timescales of the outbursts the
models require $\alpha \simeq 0.001 - 0.003$. These values are
significantly lower those required for discs undergoing physically
analogous outbursts in the binary systems, although the outbursts in
FU Ori systems seem to need to be mediated by the presence of a planet
(Lodato \& Clarke, 2004). Thus, either there is some subtlety at work
here, or the thermal disc outburst model for FU Ori outbursts does not
work. In this regard we note that other possibilities for causing FU
Ori outbursts have indeed been discussed, such as a collision of a
protostar and disc with another star (cf. Bonnell \& Bastien, 1992;
Reipurth \& Aspin, 2004).

\subsection{Summary}

We conclude that in the most clear cut cases there appears to be
strong observational evidence that values of $\alpha = 0.1 - 0.4$ are
required to provide a good description of the behaviour of
fully-ionized, thin accretion discs.

\section{Theoretical estimates from numerical simulations}

Since the work of Balbus \& Hawley (1991) there has been a large
number of publications investigating the properties of the MRI and its
relevance to viscosity in accretion discs.

Theoretical simulations of disc viscosity come in two flavours --
those which assume a superimposed seed net poloidal field, and those
which do not.  Hawley, Gammie \& Balbus (1995) showed that simulations
with a superimposed net $B_z$ tend to yield estimates of $\alpha$
larger by an order of magnitude than those which do not. They also
found that the value of $\alpha$ produced depended almost linearly on
the magnitude of the externally imposed $B_z$. Those simulations which
do not have an externally imposed $B_z$ mostly start with either a
small seed toroidal field, or alternate regions of positive and
negative vertical field $B_z$.

It seems to us an unlikely proposition that each disc is subject to a
superimposed, immovable, net poloidal field component, of exactly the
right magnitude to give rise to the $\alpha$ in the right range. For a
typical dwarf nova in outburst we find the equipartion field in the outer
disc regions to be (Shakura \& Sunyaev, 1973; Frank, King \& Raine,
2002)
\begin{eqnarray}
\lefteqn{B_{\rm eq} \simeq 5.2 \times 10^4 \left( \frac{\alpha}{0.1}
\right)^{-9/10} \left( \frac{\dot{M}}{10^{18} {\rm g/s}} \right)^{17/40}
\left( \frac{M}{M_\odot} \right)^{7/16}}\nonumber\\ 
\lefteqn{\times\left( \frac{R}{10^{10} {\rm cm}} \right)^{-21/16} {\rm G}.}
\end{eqnarray}
Here $\dot M$ is the accretion rate, $M$ the mass of the central white
dwarf, and $R$ the disc radius, with typical values assumed.  In these
numerical simulations, assumed values of $\beta_z$ (here $\beta_z = 8
\pi P / B_z^2$, where $P$ is the pressure) in the plane of the
disc are in the range 25 -- 400. This implies typical vertical fields
in the range $B_z \simeq$ 0.05 -- 0.2 $B_{\rm eq}$. For a dwarf nova
in outburst this corresponds to fields of several hundred to several
thousand Gauss. There seems no obvious source for fields of such a
magnitude on such a global scale.  Moreover, a vertical global field
threading a disc can be expected to give rise to a wind or jet from
the disc (Lovelace, Romanova \& Contopoulos, 1992; Pelletier \&
Pudritz, 1992) and a rough estimate of the mass loss of such a
jet/wind is given by Pringle (1993) as
\begin{equation}
\frac{\dot{M}_{jet}}{\dot{M}} \sim \frac{R}{H} \beta_z^{-1} \alpha^{-1}.
\end{equation}
For typical values of $\alpha \sim 0.1$ and $H/R \sim 1/30$ it is
evident that the typically assumed values of the globally imposed
field would be able to generate wind mass losses comparable to the
disc accretion rates. 

In view of all this it seems sensible to assume that the numerical
simulations most likely to correspond to physical reality are those
with no net poloidal field. In the following we consider some
representative simulations. These all use full MHD, and thus
correspond to fully ionized discs.

Early simulations were carried out by Stone et al. (1996) who consider
a shearing box, with vertical structure confined within $-2H < z <
2H.$ They adopt periodic boundary conditions in the vertical
direction, essentially for numerical reasons. In these simulations the
seed fields either have zero net vertical field or are purely
poloidal, and $\alpha$ is defined as the time and volume averaged
stress, normalized to the initial mid-plane pressure. The measured
$\alpha$ is stated as $< 0.01$ for most of the simulations. At about
the same time, Brandenburg et al. (1995) also considered a shearing
box, but now with magnetic field kept vertical at the $z$-boundaries,
though with zero net vertical flux. They found $0.001 < \alpha <
0.005$.

More recent work gives rise to similar values. For example, Sano et
al. (2004) consider a shearing box with vertical periodic boundary
conditions and no vertical gravity and examine the dependence of the
saturation level of the MRI on gas pressure. For simulations with no
net vertical flux they find $5 \times 10^{-5} < \alpha < 0.01$, with
an average value of $\alpha = 0.001$.

Miller \& Stone (2000) consider a shearing box with vertical gravity
and extend the computational domain to $\pm 5 H$ in the vertical
direction. They find that the regions away from the disc plane are
strongly magnetic, with $\beta$ as low as around 0.02, and term this
region a `corona'. This coronal region is however quite unlike the
solar corona, or stellar coronas, in that the field is quiescent
(being strong enough to stabilise the MRI) and is well ordered, being
predominantly toroidal. They find typically that $\alpha \sim 0.02$.
This work is extended by Hirose, Krolik \& Stone (2005) who study the
vertical structure of gas pressure dominated accretion discs with
local dissipation of turbulence and radiative transport. They have a
shearing box, with vertical gravity, and $-8H < z < 8H$. They find a
similar disc structure with magnetically strong regions (`coronas') at
large $|z|$, and obtain $\alpha \simeq 0.016$.

Other workers have needed to move away from the shearing box
approximation and to consider more radially extended computational
domains.  Winters et al. (2003) investigate gap formation by planets
in turbulent protostellar discs. They use full MHD and thus study
fully ionized discs. They consider a cylindrical annulus $0.25 < R/R_*
< 3.75$ and a vertical grid $-2H < z/R_* < 2H$, and before adding a
planet they establish a background flow. They do not use an initial
vertical field as this is known to produce a series of evacuated gaps
in the disc (Steinacker \& Papaloizou 2002; Nelson \& Papaloizou
2003). They use an initial seed toroidal field, and their vertical
boundary conditions are periodic. They quote a global average value of
$\alpha = 0.02$.

Nelson (2005) studies the orbital evolution of low mass protoplanets
in turbulent, magnetized discs. There is no $z$--dependence, and he
uses vertically periodic boundaries and a toroidal seed field. The
grid is cylindrical with $1 < R/R_* < 5, -0.14 < z/R_* < 0.14$. The
volume averaged stress parameter $\alpha$ is found to be $\simeq
0.005$ throughout the simulation.

A discrepant note is set by Hawley \& Krolik (2001) who find a value
of $\alpha \sim 0.1$ (their Fig 13) for radii $R/R_s \le 30$. They
perform a global simulation of a disc around a pseudo--Newtonian black
hole. They start with a torus at $R=30R_s$ (with $R_s$ being the
Schwarzschild radius). The grid is in $(R,\phi,z)$ coordinates, with
$-10 < z/R_s < 10$. The transverse magnetic field components are set
to zero outside the computational domain. The initial field is
poloidal along equal density contours, which in effect implies that
there is a net $B_z$ through the inner half torus $R/R_s \le 30$ (and
a net $B_z$ of opposite sign through the outer half). The simulations
ran for $t=1500$ (in units with $GM = R_s = 1$), which is about 7
orbits at $R/R_s=10$. Thus it seems likely that the initial global
poloidal seed field is still present throughout the computation, in
which case it is not surprising that the resulting value of $\alpha$
corresponds more closely to those found in shearing box runs which
have a superimposed seed $B_z$.

It is apparent therefore that, except perhaps for those of Hawley \&
Krolik (2001), theoretical simulations relevant to fully-ionized discs
with no imposed vertical magnetic field all produce values of $\alpha
\la 0.02$, and often considerably smaller.

\section{Theoretical limitations}

The general result in need of explanation is that for fully ionized
discs, fitting the observations appears to require $\alpha \sim 0.1 -
0.4$, whereas simulations consistently yield values which are an order
of magnitude, or more, below this value. This also implies that the
simulations have much smaller magnetic fields than are actually
present, so that disc structures and  dissipation patterns, as well as
timescales, are not being simulated correctly. This could be the reason
why simple atmosphere models are unable to fit the observed spectra of
CV discs in outburst, especially the lack of Balmer jumps, and the
ultraviolet continuum (e.g. Wade, 1984). We must therefore ask whether
the simulations are missing some essential ingredient. We consider
various possibilities in turn.

\subsection{Restrictions of scale}

Shearing box simulations miss out on low values of azimuthal
wavenumber $m$. This is because the azimuthal box size is typically
around $2 \pi H$, and so these simulations can only handle $m = 0,
R/H, 2R/H, 3R/H, ...$. Of course $R/H$ is not actually defined for the
simulations, as in effect the limit $R \rightarrow \infty$ is required
for the box to be Cartesian. But the magnetic structures which
generally emerge have structures up to and including the box size (see
Figure 16 of Hirose, Krolik and Stone 2005).

The net result of this is striking. The global 3D simulations by
Hirose, Krolik and De Villiers (2004) deal with the evolution of a
(quite thick) torus. This is close to a black hole and so not in any
sort of equilibrium. Looking at the field structure we see in Figure 6
that just above the disc (in the region called the corona) the field
is strongly azimuthal. And indeed in the body of the disc, all $m$
values are clearly present, with most of the power at {\it low} values
of $m$. Thus the main effect of a small box is to eliminate the
possibility of large--scale field structures, and thus transmission of
power in the spatial spectrum to low values of $m$.

Similar considerations apply also on radial scales. In Figure 5 of
Hirose et al (2004) we see that large--scale magnetic linkages can
occur in the radial direction. Thus shearing boxes prohibit the
generation of large--scale magnetic structures either by inverse
cascade (Pringle and Tout, 1996) or by footpoint twisting
(Lynden-Bell, 2003). It may be significant that the one computation
that looked at 3D global simulations (Hawley and Krolik, 2001) did
appear to get a larger value of $\alpha$. It is difficult to be
conclusive here, since as we remarked above the simulation probably
still contained a net poloidal field in the relevant region.

We note further that the full disc computations of Winters et
al. (2003) and of Nelson (2005) are not restricted to low $m$, but do
restrict the vertical structure (periodic boundary conditions
vertically, and Nelson has no vertical gravity). From this we can
conclude that simply allowing all azimuthal values of $m$ to be
present does not by itself solve the problem.

\subsection{Boundary conditions}

Shearing box simulations have periodic azimuthal and radial boundary
conditions. The radial one is phase--shifted to take account of the
shear, which is acceptable within the limitations discussed above,
although it should perhaps be noted that the radial force is
discontinuous at the radial boundaries. The calculations of Armitage
(1998), however, which model a radially extended disc with no vertical
structure, but with a vertically imposed field, suggest the shearing
box assumption might also have a significant effect by restricting the
scale of the field in the radial direction.

However, the vertical boundary condition poses quite a
different problem for attempts to represent the relevant physics
realistically. Usually for the shearing box this too is taken to be
periodic (implying a stack of accretion discs, rather like an old juke
box). This prevents magnetic flux from escaping.  Another approach
(e.g. Brandenburg et al., 1995) assumes that the field is kept
vertical at the boundary, and so again one cannot have flux loops
escaping. Thus in general the vertical boundary conditions serve to
suppress magnetic buoyancy and Parker--type instabilities. 

Stone et al (1996) describe early attempts to use free boundaries, and
their attendant difficulties: they write ``In principle, free
boundaries that do not inhibit outgoing waves or mass motions would be
the most appropriate for modeling an astrophysical accretion
disk. However, we have encountered numerical difficulties when strong
($\beta < 1$) highly tangled fields are advected across free
boundaries. When a strong flux loop begins to cross the boundary, the
tip is `snipped' off, releasing the two ends. Magnetic tension forces
which previously confined the loop are now unbalanced, and the ends of
the loop `snap' straight, imparting a large Lorentz force to the fluid
near the boundary. These forces can produce fluid motions that disrupt
the entire disk. Since strong, highly tangled fields are an
unavoidable consequence of the nonlinear evolution of the instability,
we have found that free outflowing boundaries cannot be used to study
the long-term evolution of disks. Instead, for most of the simulations
described in this paper we adopt periodic boundary conditions in the
vertical direction. In practice, periodic boundary conditions act much
like rigid walls in that there can be no net loss of mass or magnetic
flux through them." An attempt at circumventing this problem was made
by Miller \& Stone (2000) in order to deal with the strongly magnetic
disc regions close to the boundary. In order to reduce the limitations
of the Courant condition in regions where the Alfv\'en speed, $v_A$
becomes unacceptably large they introduced the concept of an Alfv\'en
speed limiter. This limitation is effected in practice by increasing
the momentum density of the fluid by a factor $[1 + v_A^2/c_{\rm
lim}^2]$. This implies, of course, that not all the conservation
properties of the MHD equations can be retained. Hirose et al. (2006)
face similar problems which they address by imposing a density floor
and by imposing a velocity cap of around 30 times the gas sound speed
on the disc midplane. They have an outflow boundary condition, and in
line with the comments of Stone et al., they note that it needs
careful handling to ensure stability. They also add a diffusivity in
the ghost cells, and note that the sign of the Poynting flux across
the boundary is not restricted.

Fromang and Nelson (2006) present global 3D models. Their radial
extent is a factor of 8, and their azimuthal extent an angle $\pi/4$
(thus only $m= 0, 8, 16, 24, 32, ...$ are present). Their vertical
extent is 0.3 -- 0.4 times the inner radius, with at most 25 grid
cells per vertical scale height. They have $H/R = 0.07 - 0.1$. They
use both outflow and periodic vertical boundary conditions. They
comment that the latter is less physical, but has the advantage of
preserving the total flux of the magnetic field and the vanishing of
its divergence, which is evidently difficult to ensure with the
outflow condition. For the latter they use the approaches of Miller
and Stone (2000). During the simulation the upper layers of the disc
develop very strong fields, forcing them to use an `Alfv\'{e}n speed
limiter'. This seems to indicate that flux is trapped by the boundary
conditions. Indeed they find (their Section 4) that the final states
of the magnetic corona are the same for both sets of boundary
conditions. In line with other work they find an average effective
$\alpha$ = 0.004.

\subsection{Convergence of the simulations}

Many papers on the application of MRI to accretion discs often do not
include an explicit magnetic diffusivity and so allow numerical
diffusivity (at the size of the grid cells) to provide the small scale
limit for the turbulence process. Thus the saturation level of the
turbulence (and therefore the value of $\alpha$) depends on the grid
size.  Most interesting here is the paper by Sano et al (2004) who
find that the saturation level of the MRI turbulence depends on the
gas pressure in the (shearing) box. However, since all simulations
find similar (too small) values for $\alpha$, it seems unlikely that
convergence is a major issue.

There is a possible problem here however. In hydrodynamic turbulence,
we are used to thinking that the details at the smallest scales do not
greatly affect the behaviour at large scales, where transport
properties such as the Reynolds stress are determined. But it is not
clear that this is true for MHD turbulence.  Schekochihin et
al. (2004, 2005) discuss this mainly for the interstellar medium. They
argue that the magnetic field structure created by the turbulence
depends critically on the Prandtl number (i.e. the ratio between
magnetic diffusivity and viscosity) and that this in turn affects the
saturation level of the dynamo.  There is considerable discussion of
the fact that MHD turbulence gives rise to inverse cascades, and
therefore to magnetic field structures which are much larger than the
typical driving scales.

\subsection{The breakdown of the MHD approximation}

The MHD formulation used in these simulations assumes fluid velocities
$v \ll c$ and Alfv\'en velocities $v_A \ll c$. It explicitly excludes
the displacement current and so removes the possibility of
electromagnetic waves. A good discusson of the extension of the MHD
approximation to the regime where the fluid velocities approach the
speed of light is presented by Gammie, McKinney \& T\'oth (2003). In
particular they note the distinction between behaviour in the MHD
limit when fluid density $\rho \rightarrow 0$ and in the vacuum case
$\rho = 0$. Thus the usual MHD formulation is unable to deal with
regions where the densities are low enough that the approximation of
infinite conductivity breaks down and the distinction between $\rho
\rightarrow 0$ and $\rho = 0$ becomes critical. There are severe
numerical problems implementing it in regions of large density
contrast such as the interface between the solar interior and the
solar corona, where the fields are essentially force--free.

There are also significant numerical problems dealing with
magnetically dominated outflows, or Poynting--flux dominated jets. The
problem is not simply that the timesteps get very small in numerical
simulations where the density gets small (and so $v_A$ is large) --
the point is that the MHD approximation may break down. We note that
numerical studies of the solar corona encounter these problems
(e.g. Galsgaard \& Parnell, 2005; T\"or\"ok \& Kliem, 2005; Mackay \&
van Ballegooijen, 2006a, b) and considerable complication is involved
in dealing with them.

\section{Discussion}

We have shown that there is a large discrepancy between the values of
the viscosity parameter $\alpha$ which is required to model
observations of fully ionized, time-dependent accretion discs (Section
2: $\alpha \approx 0.1 - 0.4$) and those which are generally obtained
from numerical MHD simulations without including a superimposed
magnetic field (Section 3: $\alpha \le 0.02$). In Section 4 we have
described some of the limitations inherent in the numerical
simulations and have noted that most of these would indeed tend to
lead to underestimating the value of $\alpha$. We here discuss some
other theoretical means of angular momentum transport involving global
field structures which lie outside the scope of current numerical
simulations.

We have noted in Section 4 that one of the major restrictions in the
numerical simulations is on the global structure of the magnetic
field. This suppression is driven in part by the exigency of limited
computer resources and in part by the limitations imposed by the
boundary conditions. Perhaps the closest analogy we have to what might
be happening in an accretion disc comes from looking at the behaviour
of the magnetic field at the surface of the sun. Here fluid motions
below the surface, where the MHD approximation holds reasonably well,
drive the generation of buoyant loops of magnetic flux which rise up
through the surface layers into regions where the MHD approximation is
poor, or even breaks down. These buoyant loops give rise all kinds of
complicated magnetic phenomena, including prominences, flares and the
solar wind, many of the details of which are still poorly
understood. But it is evident that the fields and flux loops extending
outside the solar surface are quite global in extent. And the radial
extent of flux loops can be much greater in stars which are rapidly
rotating (for example extending at far as 5 times the stellar radius
in AB Dor; Hussain et al., 2002). It is well known that small-scale
twisting of magnetic footpoints on the solar surface can give rise to
large-scale changes in the global structure of the field (Aly, 1984;
Sturrock, 1991), and numerical simulation of the behaviour of magnetic
fields in the low-$\beta$ regions of the solar atmosphere, driven by
motions in the high-$\beta$ regions, is an active area of research
(e.g. Galsgaard \& Parnell, 2005; T\"or\"ok \& Kliem, 2005; Mackay \&
van Ballegooijen, 2006a, b). In accretion discs, as stressed by
Ustyogova et al. (2000) and by Lynden-Bell (2003), all these phenomena
are likely to be present but driven much more vigorously by the strong
disc shear and by the fact that the disc is rotating so rapidly that
it is centrifugally supported (and so disc velocities are around 70
percent of the local escape speed).

Tout \& Pringle (1996) argue that although the dynamo process in the
disc is likely to give rise to magnetic field structures with
predominant poloidal length-scales of order $\sim H$, it is reasonable
to assume that the interaction and reconnection of such structures
outside the plane of the disc can give rise to an inverse cascade,
producing field structures of order $\sim R$ and greater. They suggest
that the large-scale poloidal fields generated in this manner can be
strong enough to power outflows and jets. Of course the presence of a
magnetically driven outflow can remove angular momentum from the disc,
and so drive accretion, even without a formal kinematic viscosity or
$\alpha$. But in neither the dwarf novae nor the SXTs is there
evidence for such outflows. However, even without driving an outflow,
large-scale poloidal fields outside the plane of the disc can provide
a non-negligible contribution to angular momentum transport, and hence
$\alpha$, of the kind that does not become apparent from, and cannot
easily be addressed in, the kind of numerical simulations discussed
above. This transport comes about for three main reasons. 

First, such a process can more easily magnetically link disparate disc
radii than processes which require radial penetration of disc material
(cf. Armitage, 1998; Fromang \& Nelson, 2006). Once linked,
differential shear winds up the field, increasing the magnetic energy
at the expense of rotational energy, and therefore transfers angular
momentum (cf. the disc-magnetosphere interaction; Livio \& Pringle,
1992).

Second, even if such large-scale poloidal flux loops do not give rise
to a steady outflow or wind, they may well give rise to intermittent
outward acceleration of disc material. As noted by Blandford \& Payne
(1982), once a magnetic field line in a centrifugally supported disc
makes an angle of greater than 60$^\circ$ to the vertical, material
can be centrifugally accelerated along it. While Blandford \& Payne
considered a constant field structure with a steady outflow, the same
physics applies equally well to intermittent field structures. Thus
one can envisage a continuous process by which small patches of disc
material are from time to time accelerated outwards for brief periods
as and when the local field configuration becomes favourable. The disc
material acquires angular momentum in the process, but not enough
energy to be expelled, and presumably falls back onto the disc at some
larger radius. Such a process leads to outwards transport of angular
momentum by a direct flux of material moving in regions out of the
disc plane. We note in passing that such a process provides a more
plausible radial transport process, required perhaps to explain the
properties of crystalline silicate grains in the pre-solar nebula,
than trying to mix material upstream in an accretion disc (e.g. Gail,
2001).

Third, the global field envisaged by Tout \& Pringle (1996) generated
by an inverse cascade from the tangled disc field is of necessity much
weaker than the mean dynamo field in the disc. But, as the numerical
experiments seem to indicate, the presence of a weak poloidal field
can serve to increase the strength of the dynamo process, and
therefore the local value of $\alpha$. Thus it may be that the disc
itself is capable of generating and sustaining, at least in a
time-averaged sense, the kind of weak global poloidal field required
to enhance the value numerically estimated of $\alpha$.

\section{Conclusion}

Over the last decade, thanks mainly to numerical simulations, we now
have a much better understanding of what is the likely driving
mechanism for accretion discs.  We have noted here that there is,
however, roughly a factor of 10 discrepancy between observational and
theoretical estimates of the accretion disc viscosity parameter
$\alpha$. We have suggested possible lines for resolving this
problem. While recognizing that this is at best close to the limits of
what is currently computationally feasible, we suggest that it is
essential to undertake fully three-dimensional, global simulations,
preferably in a large enough computational domain that the boundary
conditions have little effect on dynamics of the thin disc. We have
also noted reasons why even this may not be adequate, and have drawn
the analogy with current attempts to understand the driving of
chromospheric and coronal solar activity by subphotospheric
motions. Evidently there may be some way to go before we have a truly
predictive theory of accretion discs.

\section{Acknowledgments} 

ARK acknowledges a Royal Society--Wolfson Research Merit Award. JEP
thanks STScI for hospitality and for continued support under the
Visitors' Program. We thank the referee for helping to clarify the
contents of the paper.

\label{lastpage}

\end{document}